\documentclass[pre,twocolumn,final,showpacs,showkeys,unsortedaddress]{revtex4}
\usepackage{graphics}

\begin{document}

\title{Subwavelength imaging by a left-handed material superlens}

\author{X. S. Rao} \email{tslrx@nus.edu.sg}
\affiliation{Temasek Laboratories, National University of Singapore, Singapore 119260}

\author{C. K. Ong}
\affiliation{Centre for Superconducting and Magnetic Materials and Department of Physics, National University of Singapore, Singapore 117542}

\date{\today}

\begin{abstract}
In this work,  a finite-difference time-domain (FDTD) method is employed to justify the superlensing effect of left-handed material (LHM) slabs. Our results demonstrate that subwavelength resolution can be achieved by realistic LHM slabs with finite absorption and dimension. We present the dynamic feature of the imaging process and the dependence of physical parameters on the performance of the superlens. These results help to clarify the diversed FDTD results reported previously. 
We also show that the achievable resolution is limited by the absorption and thickness of the LHM slabs, which introduces difficulties in practical applications of the superlens.
\end{abstract}



\maketitle

\section{Introduction}
In 1968, Veselago \cite{Veselago1968} predicted that a planar slab of left-handed materials (LHMs), which possess both negative permittivity and negative permeability, could refocus the electromagnetic (EM) waves from a point source. Recently, Pendry \cite{Pendry2000} extended Veslago's analysis and further predicted that evanescent waves, which carry subwavelength structural information of the object, could be amplified inside the LHM slab and reconstructed in the image plane without loss in amplitude. Therefore, the LHM slab can be used as a {\it superlens} to achieve super-resolution, which overcomes the diffraction limit of conventional imaging systems.  
However, several recent analyses \cite{Ziolkowski2001,Valanju2002,Garcia2002,Pokrovsky2002,Loschialpo2003} contradicted Pendry's proposal. The feasibility of the superlens is still under dispute.   

There have been several attempts \cite{Ziolkowski2001,Loschialpo2003,Cummer2003} to verify the superlensing effect of a LHM slab using the finite-difference time-domain (FDTD) method. The advantage of FDTD is clear. In FDTD, the results are directly obtained from Maxwell's equations and the constitutive relations of the materials, so that unnecessary assumptions and unessential complications can be avoided. 
Ziolkowski and Heyman \cite{Ziolkowski2001} first showed that EM waves from a line source could be focused paraxially by a LHM slab, but no stable image could be formed. They found that the image moved back and forth over time and sometimes vanished altogether.  
Loschialpo {\it et al}. \cite{Loschialpo2003} 
observed a stable image of a line source formed by a LHM slab. 
They found that the image is of the order of the wavelength, showing no superlensing effect. 
However, a recent FDTD simulation by Cummer \cite{Cummer2003} reached a totally different result: subwavelength resolution of the image could be achieved by a LHM slab.
The FDTD method is supposed to be accurate and conclusive. However, these contradictory FDTD results have caused more confusions.  
Further clarifications are therefore needed.

In an earlier FDTD study of ours \cite{Rao2003}, we proposed a method to simulate different evanescent wave components separately. The method enabled us to explicitly study the dependence of wave behavior on different parameters. Our simulation results provided direct numerical evidence that evanescent waves could be amplified in a LHM slab. 
The main purpose of the present work is to extend our previous study and investigate the contribution of evanescent waves to the image quality. 
We demonstrate that super-resolution can be achieved by a LHM slab with physically realizable parameters.
Furthermore, we study the dependence of the image quality on different parameters, which may help to understand the limitation on the performance of the superlens. 
As a secondary motivation for our present work, we attempt to clarify the diversed results of previous FDTD simulations and explain why different results were obtained.

\section{Numerical method}
Following Ref.9, we consider a planer LHM slab which occupies $0\leq x\leq L$ in vacuum, with surfaces normal to $x$ direction and extends to infinity in the $y$ and $z$ directions. 
The LHM is isotropic and characterized by causal permittivity ($\epsilon$) and permeability ($\mu$) of identical plasmonic form,
\begin{equation}
\epsilon ( \omega ) = \mu ( \omega ) = 1 - \frac{\omega_p^2}{\omega^2 - i\omega \nu_c} ,
\end{equation}
where $\omega_p$ is the plasma frequency and $\nu_c$ is the collision frequency.  
A $z$-polarized electric field source is excited in the plane at $x=-L/2$ and its image is supposed to form in the plane of $x=3/2L$. The source is sinusoidal of $\omega\approx 1.1\times 10^{10}$ rad/s and with a smooth turn-on of 30-periods \cite{Ziolkowski2001}. By choosing $\omega_p$ and $\nu_c$ appropriately, we have $\epsilon (\omega)=\mu (\omega)=-1-i\gamma$ ($\gamma$ denotes the loss term) in all our simulations.
Following the method introduced in our previous work \cite{Rao2003}, we simulate the wave components with different transverse wave numbers separately by applying the boundary condition in the $y$ direction for a given $k_y$,
\begin{equation}
E_z(x,y\pm\Delta y)=E_z(x,y)e^{\mp ik_y\Delta y}.
\end{equation}
Here, we restrict our simulations to evanescent waves with $k_y>k_0$ ($k_0=\omega/c$), since it is the evanescent waves that are responsible for the subwavelength imaging.
In the simulations, the computational space is $4000\times 1$ cells with $\Delta x=\Delta y=0.3$ mm and the time step is $\Delta t=\Delta x /(2c)=0.5$ ps.  
At the working frequency, the wavelength of the evanescent waves ($\lambda_0$) is about $566 \Delta x$, which ensures the convergence of the simulation results. The LHM slab is located at the center of the space in the $x$-direction. At both ends in the $x$-direction, a 10-cell uniaxial anisotropic perfectly matched layer (UPML) \cite{Taflove2000} is added to truncate the computational space.

\section{Results and discussions}
\begin{figure}
\begin{center}
\includegraphics{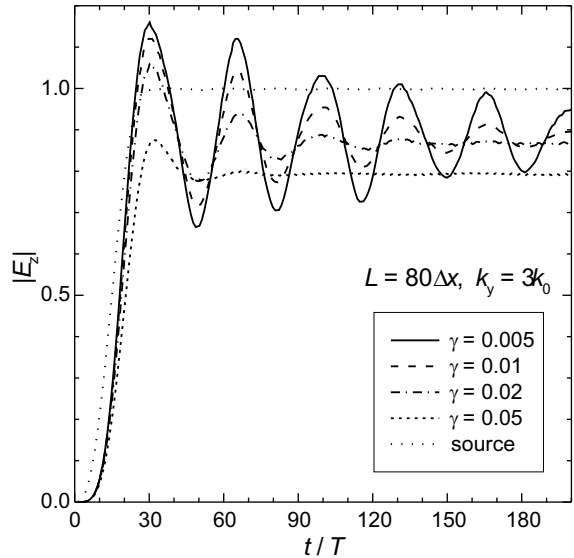}
\caption{Time evolution of the $E_z$ amplitude at the image plane for a LHM slab ($L=80\Delta x\approx 0.14\lambda_0$) with different values of $\gamma$, ranging from 0.005 to 0.05. The evanescent wave has $k_y=3.0$. The time evolution of the source field is also present for reference.}
\label{fig:1}      
\end{center}
\end{figure}
Upon stimulation, the source radiates an evanescent wave of a given $k_y$. The evanescent wave will interact with the LHM slab to form an image at the image plane. An initial transient stage is present in the simulation, and only after a certain duration can a stable image with unchanged amplitude be observed.
The time required to reach the steady state varies dramatically, and depends on different simulation parameters ($k_y$, $L$ and $\gamma$), especially the absorption $\gamma$. 
Figure 1 shows how the absorption term $\gamma$ affects the time evolution of the image. 
It is found that the transient stage, characterized by the amplitude modulation, occupies longer time for smaller values of $\gamma$. For $\gamma=0.05$, it takes less than 80 periods ($T=2\pi/\omega$) for the EM wave to reach steady state, while the time required for $\gamma=0.005$ is about 700 periods! Further reduction in $\gamma$ may require impractically long simulation time to reach the steady state. In the extreme case of zero absorption, the transient stage never decays and no steady state can be achieved. Our results are consistent with G\'omez-Santos' analysis \cite{Gomez-Santos2003}.
The dynamical feature of the imaging process explains why the FDTD simulation in Ref.3 did not show stable image. 
The authors attributed the instability of the image to the dispersive nature of the LHM. However, we believe that the unstable image comes directly from the zero or very small absorption used in their numerical examples. 

\begin{figure}
\begin{center}
\includegraphics{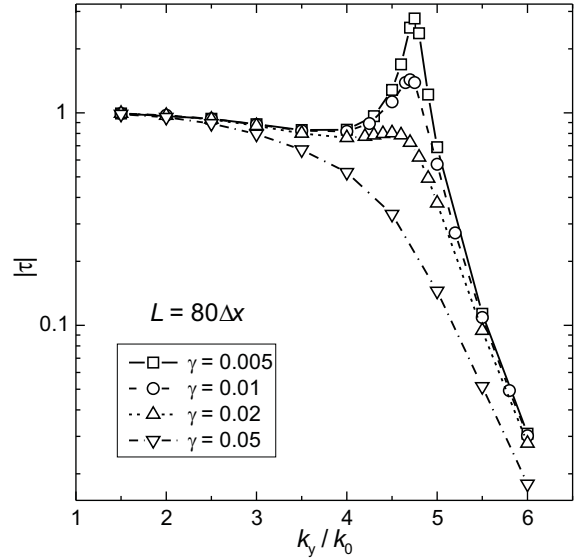}
\caption{Transfer function of a LHM slab ($L=80\Delta x\approx 0.14\lambda_0$) for different values of $\gamma$, ranging from 0.005 to 0.05, as a function of normalized transverse wave number $k_y/k_0$.}
\label{fig:2}      
\end{center}
\end{figure}
Next, we discuss the steady state results obtained. Figure 2 shows the transfer function ($\tau$), defined by the ratio of the field at the image plane to that of the source, as a function of the normalized transverse wave number $k_y/k_0$. The simulation is carried out with a LHM slab of thickness $L=80\Delta x$ for different values of absorption. 
In the case of small absorption, for example $\gamma=0.005$, the transfer function is close to unity for small $k_y$ ($|\tau|$=0.994 at $k_y=1.5k_0$) and decreases slowly with increasing $k_y$. Beyond a prominent peak ($|\tau|$=2.783) at $k_y=4.75k_0$, the transfer function decays exponentially. 
The emergence of the peak results from the resonance of the coupled surface polaritons \cite{Rao2003}. Due to finite absorption, the resonant divergence \cite{Ramakrishna2002,Smith2003} is removed. In practice, the location of the peak can be used as an estimate for the image resolution \cite{Fang2003}, which is $\sim \lambda_0/5$ for the case of $\gamma=$0.005, showing clear evidence of the superlensing effect. 
The resonance peak is also observed for $\gamma=$0.01 and 0.02. As expected, it becomes less prominent with increasing $\gamma$. Meanwhile, the peak position shifts to smaller $k_y$, though the shift is insignificant. Except for the suppression of the resonance peak, the three curves at $\gamma=$ 0.005, 0.01 and 0.02 show similar features and fit well with one another. This indicates that the superlensing effect of the LHM slab is insensitive to absorption change over certain range.
However, further increase in $\gamma$ to 0.05 not only totally suppress the resonace peak, but also cause noticeable deviation in the transfer function, especially for large $k_y$, showing degradation in the image resolution. 

\begin{figure}
\begin{center}
\includegraphics{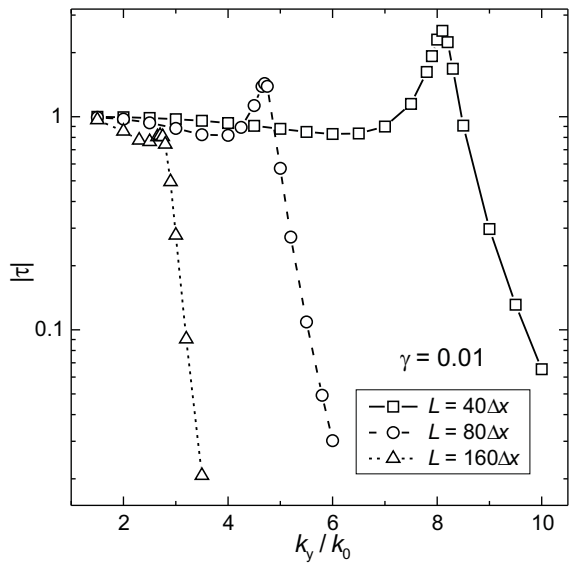}
\caption{Transfer function for LHM slabs ($\gamma=0.01$ for all) of different thicknesses as a function of normalized transverse wave number $k_y/k_0$.}
\label{fig:3}      
\end{center}
\end{figure}
We also investigate the effect of thickness of the LHM slab ($L$) on the performance of the superlens. Figure 3 shows the steady state results obtained for three LHM slabs of different thicknesses but with the same aborption. The figure clearly illustrates that $L$ has a great impact on the performance of the superlens. The resolution of the image can be dramatically improved when the thickness of the LHM slab is reduced. The image resolution for the LHM slabs of $L=40\Delta x$($\approx 0.07\lambda_0$), $80\Delta x$($\approx 0.14\lambda_0$) and $160\Delta x$($\approx 0.28\lambda_0$) are $\sim \lambda_0/8$, $\lambda_0/5$ and $\lambda_0/3$, respectively. 
It is noted that the achieved resolution is of the same order of the thickness of the slab.
It is expected that if $L$ is large enough ($\geq \lambda_0$), the resolution enhancement by the LHM slab may be totally suppressed, i.e., no evanescent waves could be reconstructed at the image with considerable amplitude. In fact, we note that this case corresponds to the numerical example studied in Ref.7, where $L\approx 3.2\lambda_0$ ($L=9.5$cm and $\lambda_0=3$cm). Based on our simulation results, it is not surprising that the image obtained in Ref.7 showed no subwavelength resolution. 

The above simulation demonstrated that subwavelength resolution can be achieved by a realistic LHM slab, which is consistent with Cummer \cite{Cummer2003}. It is worth noting that our results are obtained directly by simulating each $k_y$ component, which is different from the approach used in Ref.8 where the amplitude of different $k_y$ components at the image are obtained using spatial Fourier transformation.
Our results also revealed that the resolution is limited by the absorption as well as the thickness of the slab. The absorption is unlikely to be an obstruction for practical application of the superlens, since we believe that absorption of the order of 0.01 should not be too difficult to achieve in artifical LHMs \cite{Greegor2003}. However, the great impact of the thickness of the slab on the achievable image resolution demonstrate a stringent constraint on the most valuable application of the superlens in optics. For example, the superlens seems unlikely to be applicable in deep-submicron photolithography or nanolithography \cite{Blaikie2002}, due to the difficluty in fabricating extremely thin LHM slab with enough mechanical strength to keep the slab flat without warping.

\section{Conclusion}
To conclude, we have carried out FDTD simulations to examine the superlensing effect of the LHM slabs. Our results have justified that the resolution of the image achieved by using the LHM slab can overcome the diffraction limit of conventional imaging systems. However, the resolution of the superlens is limited by the absorption and the physical dimension of the LHM slabs.

\end{document}